\documentclass[prl,nopacs,twocolumn,preprintnumbers,notitlepage,amsmath,amssymb,superscriptaddress]{revtex4-1}

\usepackage{color}
\usepackage[pdftex]{graphicx}
\usepackage{pgffor}
\usepackage[]{pdfpages}

\makeatletter
\AtBeginDocument{\let\LS@rot\@undefined}
\makeatother

\definecolor{jerem}{rgb}{1, 0, 0}
\definecolor{jerem2}{rgb}{0, 0, 1}
\definecolor{olivier}{rgb}{0.125, 0.26, 0.07}

\newcommand{\br}[2]{\left[\frac{#1}{#2}\right]}

\usepackage{mathtools}

\usepackage{makecell}

\usepackage{enumerate}
\usepackage{mathtools}
\usepackage{stmaryrd}

\usepackage{enumitem}

\usepackage{epstopdf}

\usepackage{textcomp}
\usepackage{gensymb}
\def \equi#1{\mathrel{\mathop{\kern 0pt\sim}\limits_{#1}}}

\newcommand{\der}[0]{\text{d}}

\begin{document}
\title{Leftward, Rightward and Complete Exit Time Distributions of  Jump Processes}
\author{J. Klinger}
\affiliation{Laboratoire de Physique Th\'eorique de la Mati\`ere Condens\'ee, CNRS/Sorbonne Université, 
 4 Place Jussieu, 75005 Paris, France}
\affiliation{Laboratoire Jean Perrin, CNRS/Sorbonne Université, 
 4 Place Jussieu, 75005 Paris, France}
\author{R. Voituriez}
\affiliation{Laboratoire de Physique Th\'eorique de la Mati\`ere Condens\'ee, CNRS/Sorbonne Université, 
 4 Place Jussieu, 75005 Paris, France}
\affiliation{Laboratoire Jean Perrin, CNRS/Sorbonne Université, 
 4 Place Jussieu, 75005 Paris, France}
 \author{O. B\'enichou}
\affiliation{Laboratoire de Physique Th\'eorique de la Mati\`ere Condens\'ee, CNRS/Sorbonne Université, 
 4 Place Jussieu, 75005 Paris, France}
 

\begin{abstract}

First-passage properties of continuous stochastic processes confined in a 1--dimensional interval are well described. However,  for jump processes (discrete random walks), the characterization of the corresponding observables remains elusive, despite their relevance in various contexts. Here we derive exact asymptotic  expressions for the leftward, rightward and  complete exit time distributions from the interval $[0,x]$ for symmetric  jump processes starting from $x_0=0$, in the large $x$ and large time limit. We show that both  the leftward probability $F_{\underline{0},x}(n)$ to exit through $0$ at step $n$ and rightward  probability $F_{0,\underline{x}}(n)$ to exit through $x$ at step $n$  exhibit a universal behavior dictated by the  large distance decay of the  jump distribution parameterized by the Levy exponent $\mu$. In particular, we exhaustively describe the $n\ll x^\mu$ and $n\gg x^\mu$ limits and obtain explicit results in both  regimes. Our results finally  provide exact asymptotics for  exit time distributions of jump processes  in regimes where continuous limits do not apply.

\end{abstract}
\date{\today}

\maketitle

{\it Introduction.} In many physical systems, exit time distributions, which quantify the time taken by a random process to exit a given confining region, play a key role in understanding the relevant time scales driving the system \cite{Redner:2001a,Kampen:1992,Hughes:1995,Gardiner:2004}. Although the geometrical constraints can be defined in any dimension, the escape of random processes from the 1--dimensional interval $[0,x]$ appears as a highly recurrent and instructive physical model in a variety of fields, ranging from chemical reaction kinetics \cite{Kampen:1992,Gardiner:2004}, foraging animals \cite{Edwards2007} or financial asset modeling \cite{Kou2003,Yin2013}. A classical example of application is the Wright-Fisher evolutionary model  \cite{Wright1931}, describing the dynamics of a population of two alleles $A$ and $B$. The first time $n$ at which one of the alleles completely disappears from the population is schematically described by the first exit time distribution of a random process in the interval $[0,1]$ with initial position $x_0$ describing the initial fraction of - say - allele $A$. In fact, in this representative example of stochastic process with two alternative outcomes, not only the exit time, but also the exit side matters.  The fixation or extinction  time distributions of the allele A are indeed given respectively by the rightward or leftward exit time distributions of the corresponding process.

While these observables are well documented for 1--dimensional  continuous stochastic processes \cite{Redner:2001a,Gardiner:2004}, their discrete time counter parts, namely for jump processes, remain elusive; this is in essence because the integral equations satisfied by exit time distributions are notoriously difficult to analyze in bounded domains \cite{VanKampen1992}. Jump processes are however relevant to a variety of situations \cite{Ziff:2009vh}, and have  been the subject of multiple recent works,  in the context of self propelled particles, such as active colloids, or larger scale animals \cite{Romanczuk:2012fk,Tejedor:2012ly,Levernier:2021aa,Meyer:2021tx,Mori2020a}. In addition, experimental data of typical tracking experiments (be it of single molecules, animals or asset prices) are discrete in time by nature, because of a finite sampling rate, and  constitute intrinsic realizations of jump processes. In what follows, we focus on leftward,  rightward and complete exit time distributions of general  jump processes.

\begin{figure}[h!]
	\centering
	\includegraphics[scale=1.6]{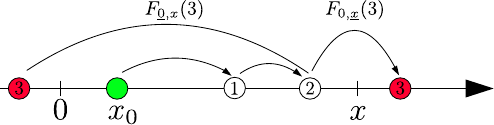}
	\caption{Rightward and leftward FETPs. In this specific realization,  after taking two steps inside the interval, the jump process escapes either through $x$ or through $0$ on its third step, with respective probabilities $F_{0,\underline{x}}(3|x_0)$ or $F_{\underline{0},x}(3|x_0)$.}
	\label{fig : 1}
\end{figure}

The $1$-dimensional jump processes considered hereafter are defined as follows : starting from $0\leq x_0\leq x$, the random walker successively performs jumps drawn from a symmetric continuous distribution $p(\ell)$, with Fourier Transform $\tilde{p}(k)=\int_{-\infty}^\infty e^{ik\ell}p(\ell)\der \ell$, until it strictly exits the interval $[0,x]$ by either crossing 0 or $x$. The corresponding  first exit time probability (FETP) at step $n$ is denoted by $F_{\underline{0},\underline{x}}(n|x_0)$. Importantly, because the random walk is defined in discrete time, the FETP is non vanishing for $x_0=0$ and thus cannot be determined by taking the continuous limit of the process, which would invariably lead to a vanishing FETP. In addition, the determination of first exit observables, and in particular the FETP for $x_0=0$,  is key in understanding  experimental data \cite{Rotter:2017uy}.  As an exemple, it was recently shown in the context of photon and neutron scattering \cite{Rotter:2017uy,Burioni:2010tr,PhysRevE.89.022135,PhysRevE.90.052114,PhysRevE.103.L010101} that the transmission probability through a slab of width $x$  was given by the splitting probability $\pi_{0,\underline{x}}$ to reach $x$ before 0 starting from 0. The latter  was determined asymptotically in \cite{Klinger2022} as
\begin{equation}\label{split_prob}
	\pi_{0,\underline{x}}\underset{x\to\infty}{\sim}\frac{2^{\mu-1}\Gamma\left(\frac{1+\mu}{2}\right)}{\sqrt{\pi}}\left[\frac{a_\mu}{x}\right]^{\frac{\mu}{2}},
\end{equation}
where $\mu$ and $a_\mu$ characterize the small $k$ behavior of $\tilde{p}(k)$:
\begin{equation}
	\tilde{p}(k)\underset{k\to 0}{=}1-(a_\mu |k| )^\mu+o(k^\mu).
\end{equation}
Of note, the splitting probability does not contain any information on the exit time. To go further and quantify the time at which exit events occur, one needs the leftward, rightward  and complete  FETPs. The leftward FETP $F_{\underline{0},x}(n)$ is defined as the probability for the walker starting from 0 to exit through 0 at time $n$ exactly without having crossed $x$ before, and $F_{0,\underline{x}}(n)$ is its rightward counterpart (see Fig. \ref{fig : 1}).
The complete FETP  $F_{\underline{0},\underline{x}}(n)$ is then given by:
\begin{equation}\label{eq : sum}
	F_{\underline{0},\underline{x}}(n)=	F_{\underline{0},x}(n)+	F_{0,\underline{x}}(n).
\end{equation}
A natural strategy to compute  these FETPs  is to consider the continuous limit of the problem, defined here as the limit $a_\mu \ll x_0$, which implies that typical exit times  satisfy $n\gg 1$.  Two limit behaviors then arise  depending on the value of $\mu$ \cite{Bouchaud:1990b,R.Metzler:2000}: for $0<\mu<2$ the process converges to an $\alpha$-stable Levy process of parameters $\mu$ and $\ a_\mu$, while for $\mu=2$ the limit distribution is that of a Brownian Motion with diffusion coefficient $D=a_2^2$. In this continuous limit, the complete first exit time distribution has been the focus of several works and is given by  \cite{Zoia:2007,Kwasnicki2012}:
\begin{equation}\label{eq : continuous_fet}
	\begin{split}
	F^{(c)}_{\underline{0},\underline{x}}(n|x_0)=\sum_{k=1}^\infty C_k(x_0)\lambda_k 2^{\mu} \frac{a_\mu^\mu}{x^\mu}e^{-\lambda_k 2^{\mu} \frac{a_\mu^\mu n}{x^\mu}}\\
	C_k(x_0)=\left[\int_0^x \psi_k\left(\frac{2}{x}u\right)\der u\right]\psi_k\left(\frac{2}{x}x_0\right)
		\end{split}
\end{equation}
where $\lambda_k$ and $\psi_k$ are respectively the eigenvalues and eigenfunctions of the fractional diffusion equation  of order $\mu$ on the interval $[0,2]$ with absorbing boundary conditions \cite{Podlubny1999}. Of note only approximates of $\psi_k$ and $\lambda_k$ have been obtained so far for $0<\mu<2$ \cite{Kwasnicki2012}. For illustration we provide $\lambda_1\simeq \left[\frac{\pi}{2}-\frac{(2-\mu)\pi}{8}\right]^\mu$; see also supplementary material (SM).

Although the continuous limit Eq. \eqref{eq : continuous_fet}  describes the regime $a_\mu \ll x_0$ for $F_{\underline{0},\underline{x}}(n|x_0)$, it fails to capture the regime $x_0 \lesssim a_\mu $ which depends on the  microscopic details of the process. In particular,  taking $x_0 \to 0$ in Eq. \eqref{eq : continuous_fet} would yield $F_{\underline{0},\underline{x}}(n|0)=0$, which is clearly incorrect for a discrete time jump process. The quantitative understanding of the regime $x_0 \lesssim a_\mu $ for leftward, rightward and complete FETPs for general jump processes, which is key to analyze experimentally relevant situations and in particular transmission properties stated above, thus calls for a new approach, which is the  objective of this paper. For the sake of simplicity, we focus here on the $x_0=0$ case (see SM for the full regime $0\leq x_0\ll a_\mu$).


{\it Summary of results.} In this letter, we derive exact asymptotics for both $F_{0,\underline{x}}(n)$ and $F_{\underline{0},x}(n)$ in the $n\to\infty$ and $x\to\infty$ limit. More precisely, we show that the rightward FETP displays the following universal asymptotic behavior:
\begin{equation}\label{eq : rep_1}
	F_{0,\underline{x}}(n) \underset{\substack{n\to\infty \\ x\to\infty \\ \tau\ {\rm fixed}}}{\sim}  \pi_{0,\underline{x}}\ h_\mu(
	\tau)\ n^{-1}
\end{equation}
where $\tau=\frac{a_\mu^\mu n}{x^\mu}$, $\pi_{0,\underline{x}}$ is the splitting probability defined above and $h_\mu$ is a universal $\mu$-dependent function. For $\mu=2$, we find \begin{equation}\label{eq : h2_summary}
	h_2(\tau) = 2\tau\pi^2\sum_{k=1}^\infty k^2(-1)^{k+1}e^{-k^2\pi^2\tau},
\end{equation}
while for $0<\mu<2$ we obtain the following asymptotic behaviors:
\begin{subequations}\label{eq : h_mu_summary}
\begin{align}
&h_\mu(\tau)\underset{\tau \ll 1}{\sim}\Gamma(\mu/2)\sin(\pi\mu/2)\pi^{-\frac{3}{2}}\sqrt{\tau}\\
&h_\mu(\tau)\underset{\tau \gg 1}{\sim} \frac{C\mu \Gamma^2(\frac{\mu}{2})}{\Gamma(\mu)}2^{\frac{\mu}{2}-2}\left[\lambda_1 2^{\mu} \tau\right] e^{-\lambda_1 2^\mu \tau}
\end{align}
\end{subequations}
where $\lambda_1$ is defined above and $C$ is a constant which reads: 
\begin{equation}\label{eq : C_def}
	C=\lim_{x_0 \to 0}\frac{C_1(x_0)}{x_0^{\frac{\mu}{2}}}.
\end{equation}

Next, we show that the leftward FETP displays an analogous universal asymptotic behavior:
\begin{equation}\label{eq : lep_1}
	F_{\underline{0},x}(n) \underset{ \substack{ n\to\infty \\ x\to\infty \\ \tau\ {\rm fixed}}}{\sim} F_{0}(n)g_\mu(\tau)
\end{equation}
where $F_0(n)\sim(4\pi n^3)^{-\frac{1}{2}}$ is the large $n$ asymptotic first passage time distribution through 0 in the semi-infinite system (starting from $0$), obtained from the celebrated Sparre-Andersen Theorem, and $g_\mu$ is a universal $\mu$-dependant function. For $\mu=2$, the function $g_2$ is determined explicitly and reads 
\begin{equation}\label{eq : g_2}
	g_2(\tau)=4 \pi^{\frac{5}{2}}\tau^{\frac{3}{2}}\sum_{k=1}^\infty e^{-k^2 \pi^2 \tau}k^2,
\end{equation}
while for $0<\mu<2$ we obtain the following asymptotic behaviors:
\begin{subequations}\label{eq : g_mu_summary}
\begin{align} 
		&g_\mu(\tau)\underset{\tau \ll 1}{\sim}1\\
		&g_\mu(\tau)\underset{\tau \gg 1}{\sim} C\,\Gamma\left(1+\frac{\mu}{2}\right)\sqrt{\pi}\lambda_1^{-\frac{1}{2}} \left[\lambda_1 2^\mu \tau \right]^{\frac{3}{2}}e^{-\lambda_1 2^\mu \tau}.
\end{align}
\end{subequations}
Finally, Eqs \eqref{eq : rep_1} to \eqref{eq : g_mu_summary} provide a comprehensive picture of the asymptotic behavior of the rightward and leftward FETPs, which in turn give access to the complete FETP.

{\it Rightward FETP.} We first write the rightward FETP as $F_{0,\underline{x}}(n) = \pi_{0,\underline{x}}\ h(x,n)$,
where $h(x,n)$ is  the conditional probability to escape through $x$ at step $n$ knowing that the walker reaches $x$ before 0, and  $\pi_{0,\underline{x}}$  is the splitting probability defined above.  In the large $n$ and $x$ limit, $h(x,n)$ can be written
\begin{equation}\label{eq : rightward_exit_continuous}
	h(x,n)=\frac{F_{0,\underline{x}}(n)}{\pi_{0,\underline{x}}} \underset{\substack{n\to\infty \\ x\to\infty\\ \tau\ {\rm fixed} }}{\sim} \lim_{x_0\to 0}\left[\frac{F^{(c)}_{0,\underline{x}}(n|x_0)}{\pi^{(c)}_{0,\underline{x}}(x_0)}\right],
\end{equation}
where $F^{(c)}_{0,\underline{x}}(n|x_0)$ is the rightward FETP of the continuous process, and $\pi^{(c)}_{0,\underline{x}}(x_0)$ the corresponding continuous splitting probability \cite{Blumenthal1961,Majumdar:2010}.

\begin{figure}[h!]
	\centering
	\includegraphics[scale=0.6]{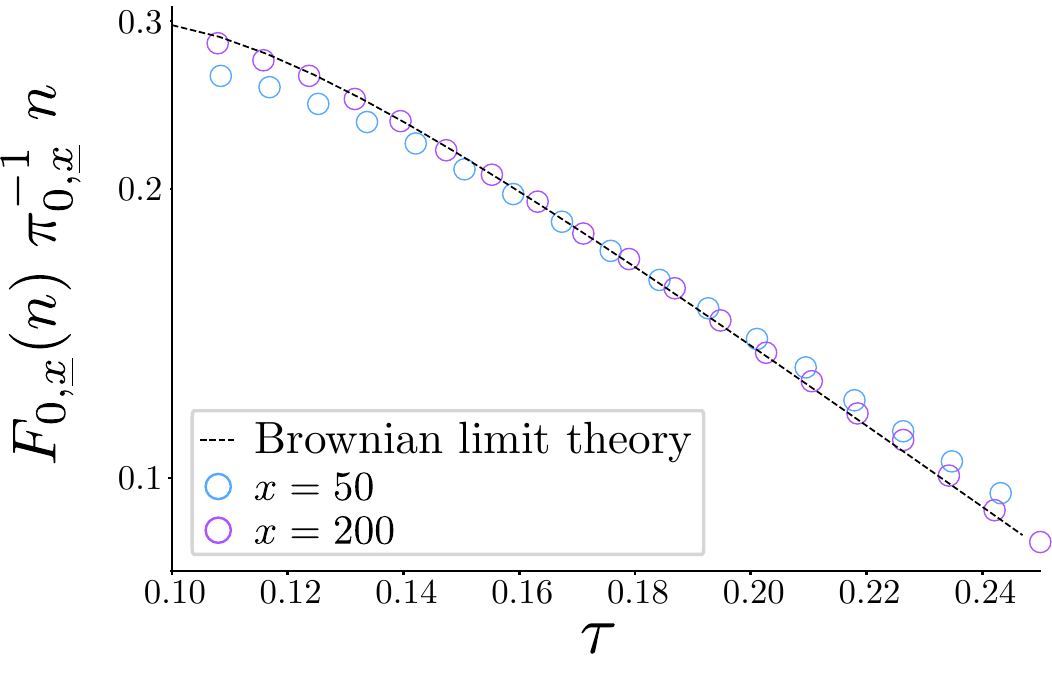}
	\caption{Rightward FETP for a jump process with $p(\ell)\propto e^{-|l|}$ (yielding $\mu=2$). Upon rescaling according to Eq. \eqref{eq : rep_1}, $F_{0,\underline{x}}(n)$ converges to the scaling function $h_2(\tau)$, defined by Eq. \eqref{eq : h2_summary}.}
	\label{fig : 2}
\end{figure}

Indeed, in the large $n$ and $x$ limit, the typical position $X_n $ of the random walker satisfies $X_n\gg a_\mu$ and the  continuous limit can be taken. In turn, since  $F^{(c)}_{0,\underline{x}}(n|x_0)\propto \pi^{(c)}_{0,\underline{x}}(x_0)$ for $x_0\to 0$ (see SM), $h(x,n)$ is a well-defined $x_0$-independent function. Making use of scale invariance, we then define the $\mu$-dependent universal scaling function  $h_{\mu}(\tau)$ -with $\tau$ given above - as: 
\begin{equation}\label{eq : h_mu_def}
		\lim_{x_0\to 0}\left[\frac{F^{(c)}_{0,\underline{x}}(n|x_0)}{\pi^{(c)}_{0,\underline{x}}(x_0)}\right]\equiv \frac{h_\mu(\tau)}{n}.
\end{equation}
This yields the result  \eqref{eq : rep_1}. Importantly, the discrete  nature of the jump process only enters through $\pi_{0,\underline{x}}$, which yields a  non vanishing rightward FETP as expected.
For $\mu=2$, $h_2(\tau)$ can be derived explicitly from Eq. \eqref{eq : rightward_exit_continuous} and leads to \eqref{eq : h2_summary}. This exact asymptotic behavior is confirmed by numerical simulations (see Fig. \ref{fig : 2}).

For $0<\mu<2$, the rightward  FETP $F^{(c)}_{0,\underline{x}}(n|x_0)$ of continuous Levy processes is not known, so that $h_\mu$ cannot be derived explicitly; its large and small $\tau$ asymptotics can however be obtained. For $\tau \gg 1$, {\it ie} $n\gg x^\mu$, we remark that the dynamics become independent of the  starting point so that $F^{(c)}_{0,\underline{x}}(n|x_0)\sim 2^{-1}F^{(c)}_{\underline{0},\underline{x}}(n|x_0)$. Using Eq. \eqref{eq : continuous_fet}, this yields  the result (\ref{eq : h_mu_summary}b). Of note, the leading $\tau$ behavior of \eqref{eq : h2_summary} is compatible with Eq. (\ref{eq : h_mu_summary}b) for $\mu=2$.

\begin{figure}[h!]
	\centering
	\includegraphics[scale=0.6]{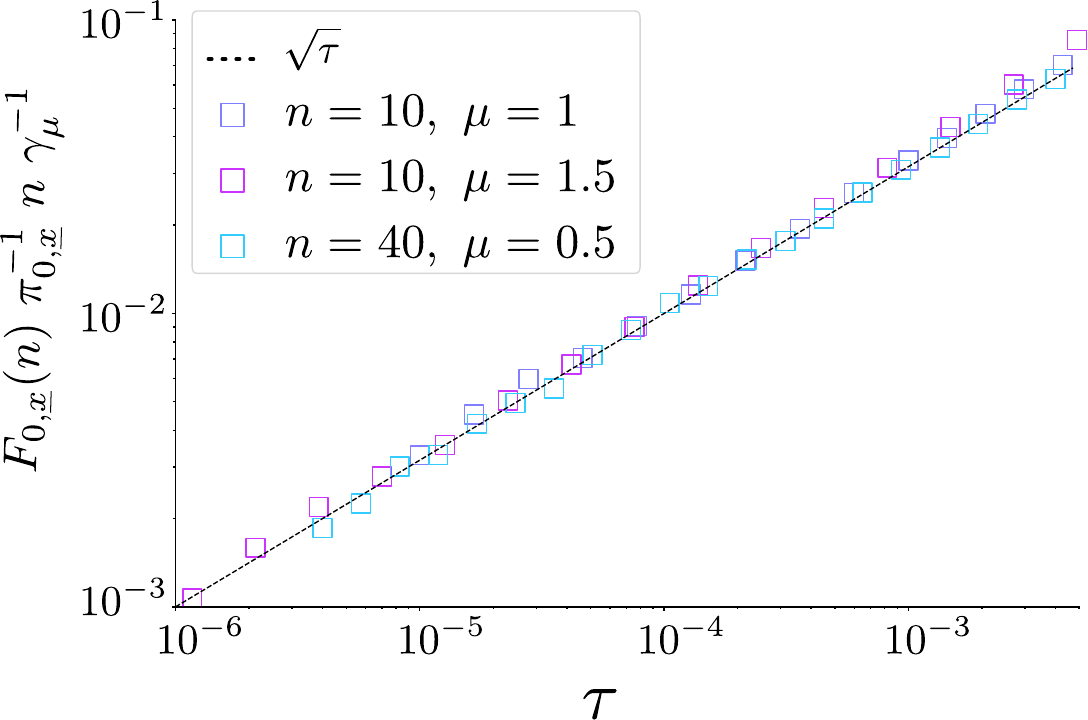}
	\caption{Small $\tau$ behavior of the rightward FETP for various Levy flights with $\mu<2$. The universal small $\tau$ behavior of $F_{0,\underline{x}}(n)$ predicted by Eq. (\ref{eq : h_mu_summary}a) is displayed, with   $\gamma_\mu=\Gamma(\mu/2)\sin(\pi\mu/2)\pi^{-\frac{3}{2}}$.}
	\label{fig : 3}
\end{figure}

For $\tau \ll 1$ (or equivalently $x\gg n^{1/\mu}$), the leading behavior of $h_\mu$ cannot be extracted from \eqref{eq : continuous_fet} because there is a priori no simple link between $h_\mu$ and $F^{(c)}_{\underline{0},\underline{x}}(n|x_0)$ in this limit. However, it can conveniently be obtained by making use of the following exact decomposition of $F_{0,\underline{x}}(n)$, which states that during the first $n-1$ steps the walker remains in the interval $[0,x]$,  while the $n^{th}$ step takes him beyond $x$:
\begin{equation}\label{eq : F_real}
	F_{0,\underline{x}}(n) = \int_0^x G_{0,x}(u,n-1)\left[\int_{x-u}^\infty p(l)\der l\right]\der u.
\end{equation}
Here $G_{0,x}(u,k)$ is defined as the propagator of the jump process in the bounded interval $[0,x]$ after $k$ steps. 
Next, we note that in the large $x$ limit with $n$ fixed, $G_{0,x}(u,n-1) \sim G_{0}(u,n-1)$ with $G_0$ the semi infinite propagator. This, together with \eqref{eq : F_real} then yields the asymptotic relation:
\begin{equation}\label{eq : F_approx}
	F_{0,\underline{x}}(n) \underset{x\to\infty}{\sim} \int_0^x G_{0}(u,n-1)U(x-u)\der u,
\end{equation}
where $U(x)=\int_x^\infty p(l)\der l$ is the cumulative of the jump distribution. Importantly this shows that  the two targets quantity $F_{0,\underline{x}}(n)$ can be expressed asymptotically in terms of the well characterized one target quantity $G_{0}(x,n)$ only. We finally introduce the Laplace transform (in space) of a given function $f(x)$ as $\tilde{f}(p)=\int_0^{\infty} e^{-px} f(x) \der x$, and the generating (function (in time) of a given function $g(n)$ as $\hat{g}(\xi)=\sum_{n\ge 0} g(n)\xi^n$ and obtain 
\begin{equation}\label{eq : F_laplace}
	\widehat{\widetilde{F}}_{0,\underline{p}}(\xi) \underset{p\to0}{\sim} \xi\  \widehat{\widetilde{G}}_{0}(p,\xi)\ \widetilde{U}(p).
\end{equation}
Both $ \widehat{\widetilde{G}}_{0}(p,\xi)$ and $\widetilde{U}(p)$ can then be readily analyzed in the $p\to0$ limit to extract the leading large $x$ behavior of $F_{0,\underline{x}}(n)$. In the case  $0<\mu<1$ (see SM for $1\leq\mu<2$), one has \cite{Ivanov1994}: 
\begin{equation}\label{eq : F_p_analysis}
	\left\{
	\begin{array}{ll} 
		& \widehat{\widetilde{G}}_{0}(p,\xi) =\frac{1}{\sqrt{1-\xi}}+o(p^\mu) \\
		&\widetilde{U}(p) = c_\mu\ a_\mu^\mu\ p^{\mu-1}+o(p^{\mu-1})
	\end{array}
	\right.
\end{equation}
where $c_\mu=\sec(\frac{\pi \mu}{2})/2$.  To leading order in $p\to 0$, we obtain $\widehat{\widetilde{F}}_{0,\underline{p}}(\xi)\sim\frac{\xi}{\sqrt{1-\xi}}c_\mu\ a_\mu^\mu\ p^{\mu-1}$ and, upon Laplace inversion, we derive the following exact asymptotic form:
\begin{equation}\label{eq : F_asymptotics}
	F_{0,\underline{x}}(n) \underset{x\rightarrow\infty}{\sim} q(n-1)\frac{\Gamma(\mu)}{\pi}\sin\Big(\frac{\pi\mu}{2}\Big)\left[\frac{a_\mu}{x}\right]^\mu
\end{equation}
where $q(n)$ is the (survival)  probability that a symmetric jump process starting from $x=0$ remains positive up to step $n$, given by the universal Sparre Andersen result  $q(n)=\binom{2n}{n}2^{-2n}$ \cite{sparre}. In fact, we show in SM that equation \eqref{eq : F_asymptotics} holds for all $\mu$ such that $0<\mu<2$. Last, using $q(n)\sim (\pi n)^{-\frac{1}{2}}$ for $n$ large,  identification with Eqs. \eqref{eq : rep_1}, \eqref{split_prob} yields the announced universal small $\tau$ behavior (\ref{eq : h_mu_summary}a), as displayed in Fig. \ref{fig : 3} for different $\mu<2$.

Of note, both asymptotic behaviors described by Eq. \eqref{eq : h_mu_summary} are necessary to recover the large $x$ scaling of the splitting probability $\pi_{0,\underline{x}}=\sum_{n=1}^\infty F_{0,\underline{x}}(n)$ (see SM).

\begin{figure}[h!]
	\centering
	\includegraphics[scale=0.58]{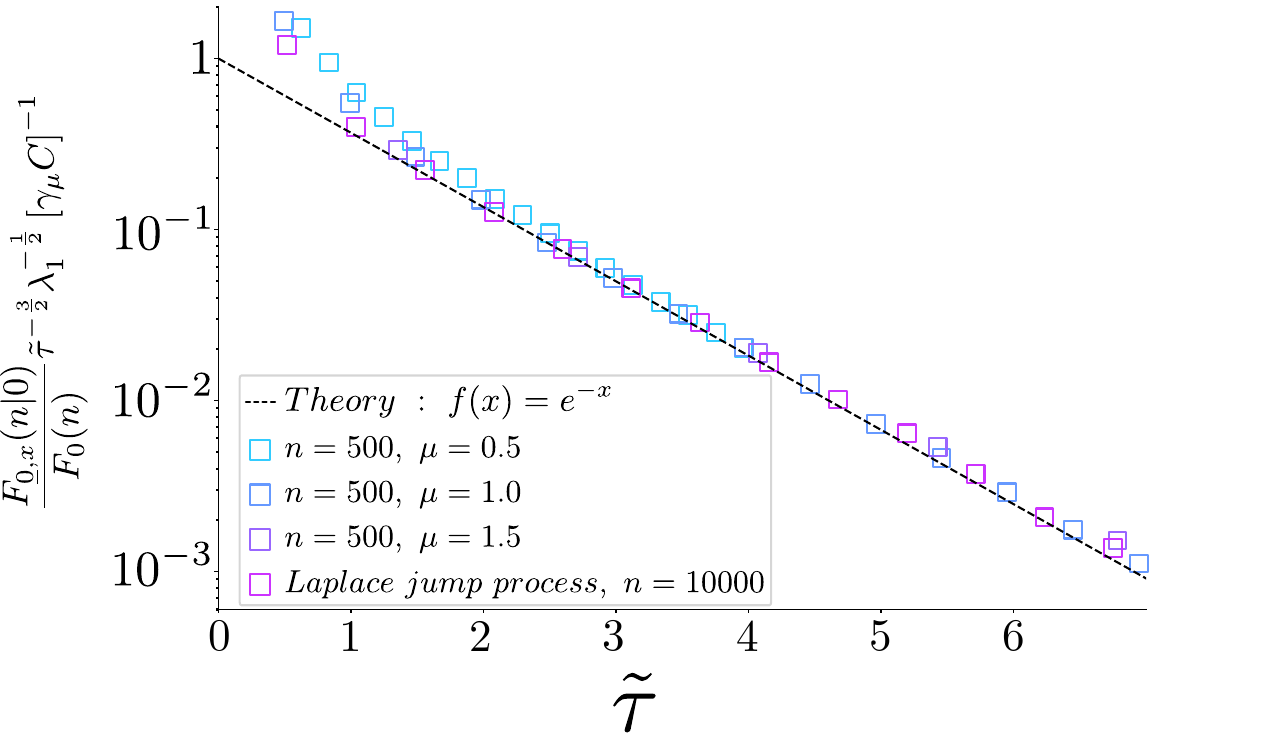}
	\caption{Leftward FETP for $n\gg x^\mu$. Defining $\tilde{\tau}=\lambda_1 2^\mu \tau$ and rescaling $F_{\underline{0},x}(n|0)$ according to (\ref{eq : g_mu_summary}b), all curves collapse onto a single exponential for various processes with $0<\mu\leq2$. The Laplace jump process is defined by $p(\ell)\propto e^{-|\ell|}$, corresponding to $\mu=2$. $C$ is given by Eq.  \eqref{eq : C_def}, and here $\gamma_\mu=\Gamma (1+\frac{\mu}{2})\sqrt{\pi}$.}
	\label{fig : 4}
\end{figure}

{\it Leftward FETP.} As for the rightward FETP, our strategy consists in expressing the two targets quantity $F_{\underline{0},x}(n)$ in terms of a well characterized one target quantity -- here the first passage time probability through 0 for a jump process starting from 0  in a semi infinite domain $F_0(n)$. We first recall that for a given jump process, the typical number of steps needed to cover a distance $x$ scales as $n\propto x^\mu$ \cite{Bouchaud:1990b}.  We thus  argue that, for an interval of typical extension $x^\mu\gg n$, $F_{\underline{0},x}(n)\sim F_0(n)$, because trajectories approaching  the rightmost target are very unlikely \cite{Levernier:2018qf,Majumdar:2010}. On the other hand, for $n\gg x^\mu$, $F_{\underline{0},x}(n)$  vanishes exponentially fast since it is increasingly unlikely for the walker to remain in $[0,x]$. Following the derivation of \eqref{eq : rightward_exit_continuous},\eqref{eq : h_mu_def}, we introduce $g(x,n)$ and define its continuous limit $g_\mu(\tau)$ by :
\begin{equation}\label{eq : g_func}
	 g(x,n)=\frac{F_{\underline{0},x}(n)}{F_{0}(n)}\underset{ \substack{ n\to\infty \\ x\to\infty \\ \tau\ {\rm fixed}}}{\sim}\lim_{x_0\to 0}\left[\frac{F^{(c)}_{\underline{0},x}(n|x_0)}{F^{(c)}_{0}(n|x_0)}\right]\equiv g_\mu(\tau)
\end{equation}
with $F^{(c)}_0$ and $F^{(c)}_{\underline{0},x}$ respectively the semi-infinite first passage time distribution and leftward FETP of the limit continuous process. It is shown in SM that  $F^{(c)}_{\underline{0},x}(n|x_0)\propto F^{(c)}_{0}(n|x_0)$ for $x_0\to 0$, which ensures that $g_\mu(\tau)$ is well defined and independent of $x_0$. 
Similarly to the rightward FETP, $g_2$ can be computed explicitly and is given in \eqref{eq : g_2}.
For  $0<\mu<2$, only the  asymptotic behavior of $g_\mu$ for $\tau\ll1$ and $\tau\gg1$ can be obtained. For small $\tau$, one has $F_{\underline{0},x}(n)\sim F_{0}(n)$ (as discussed above), yielding equation (\ref{eq : g_mu_summary}a). Note that this is verified explicitly in the case $\mu=2$  (see SM). When $\tau \gg 1$,  we perform the same analysis as for the rightward FETP. $F^{(c)}_0(n|x_0)$ is known exactly \cite{Koren2007}:
\begin{equation}\label{eq : fpt_levy}
F^{(c)}_0(n|x_0)\underset{n\to\infty}{\sim} \br{x_0}{a_\mu}^{\frac{\mu}{2}}\frac{1}{2 \sqrt{\pi} \Gamma\left(1+\frac{\mu}{2}\right)}\frac{1}{n^{\frac{3}{2}}}
\end{equation} 
and, in the large $n$ limit, $F^{(c)}_{\underline{0},x}(n|x_0)\sim 2^{-1} F^{(c)}_{\underline{0},\underline{x}}(n|x_0)$. Eq. \eqref{eq : g_func} together with Eq. \eqref{eq : continuous_fet} then yields  (\ref{eq : g_mu_summary}b), which is illustrated in Fig. \ref{fig : 4} for various $\mu\leq2$.

{\it Complete FETP.} Finally, the complete FETP can now be obtained from Eq. \eqref{eq : sum}. For $n\ll x^\mu$, one finds $F_{\underline{0},\underline{x}}(n)\sim F_{0}(n)$,  which simply reflects the fact that the target at $x$ is never approached by the walker and rightward exit events almost never occur. For $n\gg x^\mu$ however, both rightward and leftward FETP contribute and one has  $F_{0,\underline{x}}(n)\sim F_{\underline{0},x}(n)$. Indeed, after a large number of steps, the dynamics is independent of the  initial condition and exits on both sides are equiprobable. The complete FETP thus reads  $F_{\underline{0},\underline{x}}(n)\sim 2 F_{\underline{0},x}(n)$.

{\it Conclusion.} We have derived  asymptotic forms for  the rightward,   leftward and complete exit time probabilities from an interval $[0,x]$ for general jump processes starting from the edge of the domain. While such first-passage properties have been  well described for continuous stochastic processes, the case of  jump processes has so far remained  elusive, despite its relevance in various contexts. In fact, continuous limits provide only vanishing expressions for starting positions close to the edge of the domain, and are thus useless to quantify important observables such as transmission or backscattering type probabilities. These are key to analyze  experimental data, such as phase delay in neutron scattering experiments. Our approach fills this gap and provides a comprehensive  picture of exit time probabilities, which  yields asymptotically explicit  universal forms  controlled by the  large distance decay of the  jump distribution only.



\section*{Supplementary material}

# Supplementary Material for "Leftward, Rightward and Complete Exit Time Distributions of Jump Processes"

J. Klinger and O. Bénichou
*Laboratoire de Physique Théorique de la Matière Condensée,*
*CNRS, UPMC, 4 Place Jussieu, 75005 Paris, France*

R. Voituriez
*Laboratoire Jean Perrin, CNRS, UPMC, 4 Place Jussieu, 75005 Paris, France*
(Dated: December 13, 2022)

This Supplementary Material presents details of calculations:

- on the eigenvalues and eigenfunctions of the fractional Laplacian.
- on the small $x_0$ behavior of the continuous quantities.
- on the rightward exit time probability for the case $1 \leq \mu < 2$.
- on the non zero initial condition $x_0$.
- on the scaling of the splitting probability $\pi_{0,\underline{x}}$.
- on the small $\tau$ behavior of $g_2(\tau)$.

## I. EIGENVALUES AND EIGENFUNCTIONS OF THE FRACTIONNAL DIFFUSION EQUATION

In this section, we reproduce results from[3] regarding the eigenvalues and eigenfunctions of the fractional Laplacian. We emphasize that the Fractional Laplacian is an important technical tool to describe Levy-like processes, but that the eigenfunctions $\tilde{\phi}_k$ and eigenvalues $\lambda_k$ of the operator in a bounded interval are not known analytically. Here we present approximates of these quantities ; we closely follow the notations of[3]. Let $D = (-1, 1)$ be the domain of interest; we aim at finding solutions to the following equation:

$$\left(\frac{\mathrm{d}^2}{\mathrm{d}x^2}\right)^{\frac{\mu}{2}} \phi(x) = \lambda \phi(x) \quad \text{for } x \in (-1, 1) \tag{S1}$$

with $\phi(-1) = \phi(1) = 0$. Denoting $\lambda_k$ the $k$th eigenvalue, sorted in increasing order, we have:

$$\lambda_k = \left(\frac{k\pi}{2} - \frac{(2-\mu)\pi}{8}\right)^\mu + O\left(\frac{1}{k}\right). \tag{S2}$$

Approximate expressions $\psi_k$ of the corresponding eigenfunctions can be obtained by combining infinite and semi-infinite eigenfunctions of the fractional laplacian, which are known explicitly. It is found that

$$\psi_k(x) = q(-x)F_{\mu_k}(1+x) - (-1)^k q(x) F_{\mu_k}(1-x) \tag{S3}$$

with

$$\mu_k = \frac{k\pi}{2} - \frac{(2-\mu)\pi}{8} \tag{S4}$$

and $F_\lambda$ defined in the following way:

$$F_\lambda(x) = \sin\left(\lambda x + \frac{(2-\mu)\pi}{8}\right) - G(\lambda x). \tag{S5}$$

Here $G$ is the laplace transform of a positive function $\gamma(s)$:

$$\gamma(s) = \frac{\sqrt{2\mu}\sin\left(\frac{\mu\pi}{2}\right)}{2\pi} \frac{s^\mu}{1 + s^{2\mu} - 2s^\mu \cos\left(\frac{\mu\pi}{2}\right)} \exp\left[\frac{1}{\pi}\int_0^\infty \frac{\mathrm{d}r}{1+r^2}\ln\left(\frac{1 - r^\mu s^\mu}{1 - r^2 s^2}\right)\right]. \tag{S6}$$

Finally $q$ is an interpolating function:

$$q(x) = \begin{cases} 0 & \text{for } x \in (-\infty, -\frac{1}{3}) \\ \frac{9}{2}\left(x + \frac{1}{3}\right)^2 & \text{for } x \in (-\frac{1}{3}, 0) \\ 1 - \frac{9}{2}\left(x - \frac{1}{3}\right)^2 & \text{for } x \in (0, \frac{1}{3}) \\ 1 & \text{for } x \in (\frac{1}{3}, \infty) \end{cases} \tag{S7}$$

From these combined expressions, one can numerically evaluate the approximates $\psi_k$; this is used to determine numerically $C$ in equation (8) of the main text.

## II. SMALL $x_0$ BEHAVIOR OF THE CONTINUOUS QUANTITIES

In this section, we show that all continuous quantities considered in the main text, namely $F_0^{(c)}(n|x_0)$, $F_{0,\underline{x}}^{(c)}(n|x_0)$, $F_{\underline{0},x}^{(c)}(n|x_0)$, $\pi_{\underline{0},x}^{(c)}(x_0)$, vanish similarly as $x_0 \to 0$. Denoting here $\psi_k$ the eigenfunctions of the fractional Laplacian operator on $[0, x]$ and $\phi_k$ the eigenfunctions of the fractional Laplacian operator on $[0, +\infty]$, Kwasnicki shows that for $x_0 \to 0$:

$$\psi_k(x_0) \propto \phi_k(x_0) \propto \left[\sqrt{\frac{\mu}{2}}\Gamma\left(\frac{\mu}{2}\right)\right]^{-1}[\mu_k x_0]^{\frac{\mu}{2}}. \tag{S8}$$

Since all of the $F_0^{(c)}(n|x_0)$, $F_{0,\underline{x}}^{(c)}(n|x_0)$, $F_{\underline{0},x}^{(c)}(n|x_0)$, $\pi_{\underline{0},x}^{(c)}(x_0)$ can be projected either on the $\psi_k$ or $\phi_k$ basis, we obtain that they all vanish as $x_0^{\frac{\mu}{2}}$ as $x_0 \to 0$.

## III. DETAILS ON THE RIGHTWARD EXIT TIME PROBABILITY FOR THE CASE $1 \leq \mu < 2$

In this section, we focus on the asymptotic behavior of $F_{0,\underline{x}}(n|0)$ in the case $1 \leq \mu < 2$, by performing the same asymptotic expansion as in the main text for the $\mu < 1$ case.

### A. $\mu = 1$ case

Let us list the various small $p$ expansions necessary for the analysis, which can be found in[4], or easily derived:

$$\begin{aligned} \widehat{\widetilde{G}}_0(p,\xi|0) &= \frac{1}{\sqrt{1-\xi}}\left[1 + \left(\frac{\xi}{1-\xi}\right)^{\frac{1}{\mu}} \frac{a_1}{\pi} p\log(p) + O(p)\right] \\ \widetilde{f}(p) &= \frac{1}{2} + \frac{a_1}{\pi} p\log(p) + O(p) \\ \widetilde{F}(p) &= -\frac{a_1}{\pi}\log(p) + O(1) \end{aligned} \tag{S9}$$

To lowest order in $p$ one then obtains:

$$\widehat{\widetilde{F}}_{0,\underline{p}}(\xi) = -\frac{\xi}{\sqrt{1-\xi}}\frac{a_1}{\pi}\log(p) + o(\log(p)) \tag{S10}$$

which yields, after inversion:

$$F_{0,\underline{x}}(n|0) \underset{x\to\infty}{\sim} \frac{1}{\pi} q(n-1|0) \left[\frac{a_1}{x}\right] \tag{S11}$$

in agreement with equation (18) from the main text.

### B.  $1 < \mu < 2$ case

Let us repeat this operation:

$$\begin{aligned}
\widehat{\widetilde{G}}_0(p,\xi|0) &= \frac{1}{\sqrt{1-\xi}} \left[1 + \left(\frac{\xi}{1-\xi}\right)^{\frac{1}{\mu}} \frac{a_\mu\, p}{\sin(\pi/\mu)} - \frac{\xi}{1-\xi} c_\mu (a_\mu p)^\mu + o(p^\mu)\right] \\
\widetilde{f}(p) &= \frac{1}{2} - p\langle f \rangle - c_\mu (a_\mu p)^\mu + o(p^\mu) \\
\widetilde{F}(p) &= \langle f \rangle + c_\mu\, a_\mu^\mu\, p^{\mu-1} + o(p^{\mu-1})
\end{aligned} \tag{S12}$$

To lowest order in $p$ one then obtains:

$$\widehat{\widetilde{F}}_{0,p}(\xi|0) = \frac{\xi}{\sqrt{1-\xi}} \left[\langle f \rangle + c_\mu\, a_\mu^\mu\, p^{\mu-1}\right] \tag{S13}$$

which yields, after inversion:

$$F_{0,\underline{x}}(n|0) \underset{x\to\infty}{\sim} q(n-1|0) \left[\langle f \rangle \delta(x) + \frac{1}{\pi}\Gamma(\mu)\sin(\frac{\pi\mu}{2}) \left[\frac{a_\mu}{x}\right]^\mu\right]. \tag{S14}$$

In the large $x$ limit, we finally obtain:

$$F_{0,\underline{x}}(n|0) \underset{x\to\infty}{\sim} q(n-1|0) \frac{1}{\pi}\Gamma(\mu)\sin(\frac{\pi\mu}{2}) \left[\frac{a_\mu}{x}\right]^\mu \tag{S15}$$

in agreement with equation (18) of the main text.

## IV.  NON ZERO INITIAL CONDITIONS

In this section we focus on the derivation of $F_{0,\underline{x}}(n|x_0)$ and $F_{\underline{0},x}(n|x_0)$ for $0 < x_0 \ll a_\mu$. Note that the assumption $x_0 \ll a_\mu$ is important since in the opposite limit $x_0 \gg a_\mu$, the continuous limit is recovered and the rightward and leftward exit time distributions are known. Our objective here is thus to highlight the peculiar behaviors arising from the discrete time nature of the jump process. Recalling equations (5) and (9) of the main text, we argue that the dependence on the initial position of exit time probabilities is contained either in the splitting probability (rightward FETP) or in the semi infinite first passage time probability (leftward FETP):

$$\begin{aligned}
F_{0,\underline{x}}(n|x_0) &\underset{\substack{n\to\infty \\ x\to\infty \\ \tau\text{ fixed}}}{\sim} \pi_{0,\underline{x}}(x_0) h_\mu(\tau) n^{-1} \\
F_{\underline{0},x}(n|x_0) &\underset{\substack{n\to\infty \\ x\to\infty \\ \tau\text{ fixed}}}{\sim} F_0(n|x_0) g_\mu(\tau)
\end{aligned} \tag{S16}$$

It was shown in[1,4] that both splitting and semi infinite first passage time probabilities exhibit similar asymptotic forms in the large $n$ and $x$ limit. More precisely, for a jump process with jump distribution $p(l)$ and Fourier Transform $\tilde{p}(k) = \int_{-\infty}^\infty e^{ik\ell} p(\ell) d\ell$ one has:

$$F_0(n|x_0) \underset{n\to\infty}{\sim} \frac{1}{\sqrt{4n^3}}\left[\frac{1}{\sqrt{\pi}} + V(x_0)\right] \tag{S17}$$

$$\pi_{0,\underline{x}}(x_0) \underset{x\to\infty}{\sim} 2^{\mu-1}\Gamma\left(\frac{1+\mu}{2}\right)\left[\frac{a_\mu}{x}\right]^{\frac{\mu}{2}}\left[\frac{1}{\sqrt{\pi}} + V(x_0)\right]$$

where $a_\mu$ is the scale of the jump process defined by the small $k$ expansion of $\tilde{p}(k)$:

$$\tilde{p}(k) \underset{k\to 0}{=} 1 - (a_\mu |k|)^\mu + o(k^\mu) \tag{S18}$$

and $V(x_0)$ is defined by its laplace transform:

$$\mathcal{L}V(\lambda) = \int_0^\infty V(x_0) e^{-\lambda x_0} dx_0$$
$$= \frac{1}{\lambda\sqrt{\pi}}\left(\exp\left[-\frac{\lambda}{\pi}\int_0^\infty \frac{dk}{k^2+\lambda^2}\ln(1-\tilde{p}(k))\right] - 1\right) \tag{S19}$$

Combining equations (S16) and (S17), we thus obtain the small $x_0 \ll a_\mu$ behavior of the rightward and leftward FETPs, along with the complete exit time distribution $F_{\underline{0},\underline{x}}(n|x_0) = F_{0,\underline{x}}(n|x_0) + F_{\underline{0},x}(n|x_0)$. Agreement with numerical simulations is displayed in figure 1.

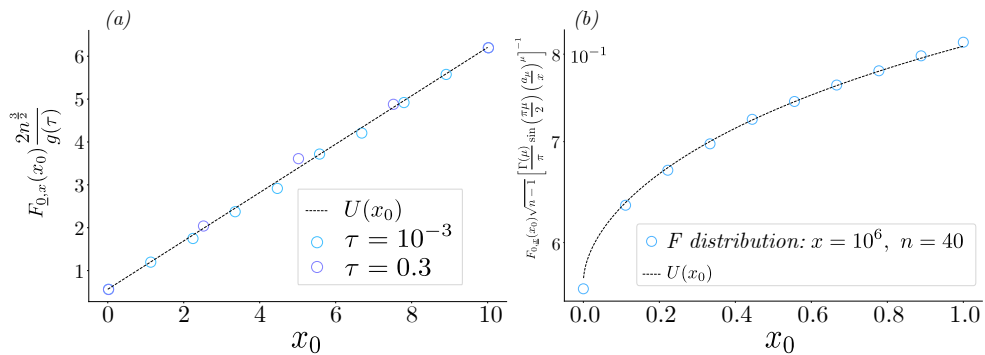

Figure 1: $x_0$ dependence of exit time probabilities. (a) Leftward FETP for a Laplace jump process. After rescaling, the rightward FETP collapses to the $U(x_0) = \frac{1}{\sqrt{\pi}} + V(x_0)$ function, for various $\tau$ values. (b) Rescaled rightward FETP for a F-distributed jump process defined by $p(\ell) \propto \left[\sqrt{|l|}(1+|l|)\right]^{-1}$. The $x_0$ dependence is in this case sub linear.

These expressions constitute the extension of the results derived in the main text to the full regime $0 < x_0 \ll a_\mu$.

## V. SCALING OF THE SPLITTING PROBABILITY

In this section, we show how to recover the scaling of the splitting probability derived in[2] from equations (5) and (18). Recall first that the splitting probability starting from $x_0 = 0$ reads:

$$\pi_{0,\underline{x}} = \sum_{k=1}^\infty F_{0,\underline{x}}(k). \tag{S20}$$

Splitting this sum into two parts we obtain the following decomposition:

$$\pi_{0,\underline{x}} = \sum_{k=1}^{x^\mu} q(k-1)\frac{\Gamma(\mu)}{\pi}\sin\left(\frac{\pi\mu}{2}\right)\left[\frac{a_\mu}{x}\right]^\mu + \sum_{k=x^\mu}^{\infty} \pi_{0,\underline{x}}\frac{h_\mu(\frac{k}{x^\mu})}{k} \tag{S21}$$
$$\equiv A_1(x) + A_2(x).$$

Since the survival probability satisfies $q(k) \propto k^{-\frac{1}{2}}$ in the large $k$ limit, one has $A_1(x) \propto x^{-\frac{\mu}{2}}$. In the large $x$ limit, the second sum can be rewritten as an integral:

$$A_2(x) \sim \pi_{0,\underline{x}} \int_1^\infty h_\mu(\tau)\tau^{-1} d\tau. \tag{S22}$$

Recalling the definition of $h_\mu(\tau)$ from equation (3) of the main text, we rewrite $A_2$ as:

$$A_2(x) \sim \pi_{0,\underline{x}} \sum_{k=1}^{\infty} P(k)e^{-\lambda_k 2^\mu} \tag{S23}$$

with $P(k)$ some sub exponential function of $k$ that guarantees the convergence of the sum. Summing $A_1$ and $A_2$ together, we thus recover the expected scaling of the splitting probability derived in[2]:

$$A_1(x) + A_2(x) \propto x^{-\frac{\mu}{2}} \tag{S24}$$

## VI. SMALL $\tau$ BEHAVIOR OF $g_2(\tau)$

In this section we give an alternate expression of $g_2(\tau)$ defined in equation (10) of the main text, which is convenient to analyze small $\tau$ values. We obtain this expression by using the Poisson Sum Formula:

$$g_2(\tau) = 4\pi^{\frac{5}{2}}\tau^{\frac{3}{2}}\sum_{k=1}^{\infty} e^{-k^2\pi^2\tau} k^2$$
$$= 1 + \sum_{k=1}^{\infty} e^{-\frac{k^2}{\tau}}\left(2 - \frac{4k^2}{\tau}\right) \tag{S25}$$

In particular, this yields $g_2(\tau) \to 1$ for $\tau \to 0$.

---



\end{document}